\documentclass{aastex631}

\usepackage[whole]{bxcjkjatype}

\begin{document}

\title{Extended and Compact Ortho-H$_2$D$^+$ Structures Close to the Moment of Star-Formation: Evidence from ALMA-ACA Observations in Taurus}

\author[0000-0002-2062-1600]{Kazuki Tokuda}
\affiliation{Faculty of Education, Kagawa University, Saiwai-cho 1-1, Takamatsu, Kagawa 760-8522, Japan}
\affiliation{Department of Earth and Planetary Sciences, Faculty of Science, Kyushu University, Nishi-ku, Fukuoka 819-0395, Japan}
\affiliation{National Astronomical Observatory of Japan, National Institutes of Natural Sciences, 2-21-1 Osawa, Mitaka, Tokyo 181-8588, Japan}

\author[0000-0002-2026-8157]{Kenji Furuya}
\affiliation{RIKEN Pioneering Research Institute, 2-1 Hirosawa, Wako-shi, Saitama 351-0198, Japan}
\affiliation{Department of Astronomy, Graduate School of Science, University of Tokyo, Tokyo 113-0033, Japan}

\author[0009-0005-4458-2908]{Naofumi Fukaya}
\affiliation{Department of Physics, Nagoya University, Furo-cho, Chikusa-ku, Nagoya 464-8601, Japan}

\author[0000-0002-1411-5410]{Kengo Tachihara}
\affiliation{Department of Physics, Nagoya University, Furo-cho, Chikusa-ku, Nagoya 464-8601, Japan}

\author[0000-0002-8149-8546]{Ken'ichi Tatematsu}
\affiliation{Nobeyama Radio Observatory, National Astronomical Observatory of Japan, National Institutes of Natural Sciences, 462-2 Nobeyama, Minamimaki, Minamisaku, Nagano 384-1305, Japan}
\affiliation{Astronomical Science Program, Graduate Institute for Advanced Studies, SOKENDAI, 2-21-1 Osawa, Mitaka, Tokyo 181-8588, Japan}

\author[0000-0003-4271-4901]{Shingo Nozaki}
\affiliation{Department of Earth and Planetary Sciences, Faculty of Science, Kyushu University, Nishi-ku, Fukuoka 819-0395, Japan}

\author[0000-0002-3297-4497]{Nami Sakai}
\affiliation{The Institute of Physical and Chemical Research (RIKEN), 2-1 Hirosawa, Wako-shi, Saitama 351-0198, Japan}

\author[0000-0003-3283-6884]{Yuri Aikawa}
\affiliation{Department of Astronomy, Graduate School of Science, University of Tokyo, Tokyo 113-0033, Japan}

\author[0000-0003-1549-6435]{Kazuya Saigo}
\affiliation{Graduate School of Science and Engineering, Kagoshima University, 1-21-40 Korimoto, Kagoshima City, Kagoshima 890-0065, Japan}

\author[0000-0001-7826-3837]{Toshikazu Onishi}
\affiliation{Department of Physics, Graduate School of Science, Osaka Metropolitan University, 3-3-138 Sugimoto, Sumiyoshi-ku, Osaka 558-8585, Japan}

\author[0000-0002-0963-0872]{Masahiro N. Machida}
\affiliation{Department of Earth and Planetary Sciences, Faculty of Science, Kyushu University, Nishi-ku, Fukuoka 819-0395, Japan}

\begin{abstract}
 
Observing and characterizing pre- and protostellar cores in the earliest and densest stages of star formation is challenging due to their short timescales and high densities, limiting the suitable tracers and targets. We conducted ALMA\UTF{2013}Atacama Compact Array (ACA) stand-alone observations of ortho-H$_2$D$^+$~(1$_{\rm 1,0}$--1$_{\rm 1,1}$) emission, which is believed to trace cold high-density regions, toward three dense cores in the Taurus molecular cloud: (1) L1544, likely in the densest prestellar phase; (2) MC~35-mm, a candidate for the first hydrostatic core; and (3) MC~27/L1521F, which hosts a Class~0 very-low luminosity object. 
These observations provide high angular resolution data for the line across a set of cores selected to represent consecutive stages around the onset of star formation, offering a unique opportunity to trace the time evolution of $\sim$10$^4$ years.
With the single-dish total-power array, we detected ortho-H$_2$D$^+$ emission in all three cores, revealing its presence over scales of $\sim$10,000 au.  In the interferometric 7\,m array data with a beam size of 3\farcs5 ($\sim$500\,au), emission was detected only toward the central continuum source of MC~35-mm, with a significance of $\sim$3$\sigma$. No significant detections were found in the other targets, placing an upper limit on the H$_2$D$^{+}$ abundance of $\sim$10$^{-11}$ in the dense components traced by the interferometric continuum emission. These results suggest that ortho-H$_2$D$^+$ predominantly exhibits an extended distribution over several thousand au in the early stages of star formation. Detection in compact, dense central structures may only be achieved within a few $\times$ 10$^{4}$ years immediately before or after protostar formation.

\end{abstract}

\keywords{Star formation (1569); Protostars (1302); Molecular clouds (1072); Interstellar medium (847); Interstellar molecules (849); Collapsing clouds (267)}

\section{Introduction} \label{sec:intro}

Understanding the formation of low-mass stars, which represent the fundamental building blocks of the universe and account for the majority of the stellar mass distribution, has a long history of research \citep[e.g.,][]{Larson_1969} and occupies a foundational position in astronomy. Dense cloud cores \cite[e.g.][]{Myers_1983,Bergin_2007}, particularly those gravitationally bound and considered to be in the phase just prior to star formation, are referred to as $``$prestellar (or pre protostellar) cores$"$ \cite[e.g.,][]{Ward-Thompsoon_1994}. Typically, they are deeply embedded within molecular clouds and lack infrared signatures, making their discovery and characterization reliant on unbiased surveys with ground-based single-dish radio telescopes \cite[e.g.,][]{Mizuno_1994,Onishi_2002} and far-infrared satellite observations \cite[e.g.,][]{Konyves_2015}. However, as one approaches the moment of star formation, observational characterization becomes increasingly challenging, and their full nature has yet to be revealed. Higher densities of prestellar cores indicate shorter free-fall timescales, resulting in shorter lifetimes in such dense states \cite[e.g.,][]{Onishi_2002,Tokuda_2020Tau}, thereby limiting the number of observable targets on the verge of star formation. Additionally, at higher (column) densities, issues such as optical thickness and molecular freeze-out onto dust grains become problematic \cite[e.g.,][]{Caselli_1999,Tatematsu_2004}, restricting the molecular species capable of tracing extremely dense regions with an H$_2$ number density of $>$10$^6$--10$^7$\,cm\,$^{-3}$.

To address these challenges by high-resolution, high-sensitivity observations with the Atacama Large Millimeter/submillimeter Array (ALMA), we selected three targets from the Taurus molecular cloud, which is one of the closest, $D$ $\sim$130--170\,pc \citep{Galli_2018,Galli_2019}, and most well-studied low-mass star-forming regions. These targets allowed us to trace the evolution of dense core centers over timescales of tens of thousands of years, capturing snapshots spanning the moment of star formation (see Section~\ref{sec:obs}). The main tracer employed in this study is ortho-H$_2$D$^+$(1$_{\rm 1,0}$--1$_{\rm 1,1}$), whose rest frequency is 372.42134\,GHz \citep{Amano_2005}. Theoretical and observational studies demonstrated that the deuterated species of H$_3^+$ are powerful tracers of core nuclei \cite[e.g.,][]{Caselli_2003,Walmsley_2004,Aikawa_2005,Bergin_2007}, even when other high-density tracers are depleted onto dust grains. However, a limitation of this tracer is that its frequency lies in a range of the spectrum where atmospheric absorption is significant, resulting in high observational noise levels. Consequently, achieving high-sensitivity observations across multiple targets is challenging. Therefore, it is crucial for observers to carefully optimize target selection, field configuration, and array settings in advance to maximize the observation efficiency.

In the ALMA era, early science observations and the subsequent follow-up program toward Oph-SMN1 \citep{Friesen_2014,Friesen_2024}, which is at a comparable distance to Taurus, achieved a successful detection in ortho-H$_2$D$^{+}$. They identified a compact emission of $\sim$500\,au associated with the first core candidate, based on combined data from the 12\,m and 7\,m arrays. While this was a seminal study, Oph-SMN1 lies within a clustered star-forming region. Therefore, it remains essential to investigate how ortho-H$_2$D$^{+}$ emission behaves in isolated star-forming cores. Our archival data investigation (2016.1.00240.S) as preliminary research confirmed that the lack of compact emission with angular sizes less than 1$\arcsec$($\sim$170\,au) in the Band~7 (0.8\,mm) continuum toward the prestellar core L1544, one of our present targets. It means that observations using the 12\,m array are not necessarily the ideal option due to spatial filtering effects. We thus conclude that Band~7 ortho-H$_2$D$^+$ observations using the Atacama Compact Array (ACA) stand-alone mode provide the best approach for obtaining an initial glimpse into these structures whose angular scale is less than a few arcseconds ($<$1,000\,au).

The immediate goal of this study is to characterize the physical conditions at the centers of dense cores in stages preceding star formation. As a first step, we experimentally investigate whether one of the deuterated isotopologues of H$_3^+$, orth-H$_2$D$^{+}$, can effectively serve as tracers of such dense core centers in the low-mass star formation regime (c.f., high-mass one by e.g., \citealt{Redaelli_2021}).

\section{Target Descriptions, ALMA Observations, and Data Reductions} \label{sec:obs}

We selected three target sources: L1544, MC35-mm, and MC~27/L1527F. The prestellar core L1544 is one of the most well-known prestellar cores. \cite{Caselli_2019} successfully detected a compact and dense structure with an angular scale of a few arcseconds ($<$1,000 au) at 1.3\,mm continuum with ALMA. They confirmed that the central density is close to $\sim$10$^7$\,cm$^{-3}$ and demonstrated that there is no protostellar object within it. MC35-mm is a recently identified candidate for the first hydrostatic core, discovered through an unbiased ACA survey in Taurus \citep{Fujishiro_2020, Tokuda_2020Tau}. This source does not host any bright ($>$0.01\,$L_{\odot}$) infrared sources but exhibits a low-velocity ($\sim$5\,km\,s$^{-1}$) molecular outflow close to the compact millimeter source, which are in agreement with the theoretical predictions in the first hydrostatic core phase \cite[e.g.,][]{Machida_2008}. MC~27/L1521F was initially classified as a prestellar core \citep[e.g.,][]{Mizuno_1994,Onishi_1999,Tatematsu_2004,Crapsi_2004} before Spitzer discovered a very-low-luminosity Class~0 protostar \citep{Bourke_2006} at its center. ALMA observations revealed that the embedded protostar has a mass of $\sim$0.2\,$M_{\odot}$ \citep{Tokuda_2017}, making it the least massive protostar in Taurus so far. 
The surrounding envelope gas exhibits a variety of complex structures, such as arcs, filaments, and spikes, spanning a wide dynamic range from a few au to several thousand au. These features are possibly formed by interactions between protostars/disks and the surrounding gas with magnetic fields \citep{Tokuda_2014,Tokuda_2018,Tokuda_2024}.

Single-dish millimeter and submillimeter observations in previous studies show that the peak dust continuum intensities of the three dense cores are nearly identical \cite[e.g.,][]{Ward_1999,Motte_2001,
Kirk_2005,Kauffmann_2008}. The H$_2$ column densities ($N_{\rm H_2}$) are estimated to be approximately 1 $\times$\,10$^{23}$\,cm$^{-2}$ for all sources at $\sim$10$\arcsec$--15$\arcsec$ beam size. Table~\ref{tab:target} summarizes the evolutionary stages of the cores, including central densities and relative timescales from protostar formation. For L1544, given its extremely high central density of $\sim$10$^{7}$\,cm$^{-3}$ and possible collapsing evidence \citep{Tafalla_1998}, the timescale until the protostar formation is expected to be represented by the free-fall time at that density \citep{Tokuda_2020Tau}. Because MC~35-mm hosts a compact source close to the first core phase, its central density within the inner $\sim$1,000 au is likely higher than that of L1544. Table~\ref{tab:target} lists the central densities and ages of the three targets. In summary, our selected combination enables comparisons on timescales of tens of thousands of years, capturing snapshots across the moment of star formation.

\begin{table}[htbp]
\centering
\caption{Target Dense Cores}
\label{tab:target}
\begin{tabular}{cccclll}
\hline \hline
Core Name & $D^{1}$ (pc) & $n({\rm H_2})$ (cm$^{-3}$) & Age$^{2}$ (yr)            & Evidence for age estimate          & References               & Other name            \\
\hline
L1544     & 171.7 & $\sim$10$^{7}$              & $\sim-$10$^4$                    & Free-fall timescale          & \cite{Caselli_2019}      & $\cdots$ \\
MC~35-mm   & 128.7 & $>$10$^{7}$                 & 0--10$^4$                 & Outflow dynamical timescale        & \cite{Fujishiro_2020}    & L1535-NE              \\
MC~27      & 140 & $>$10$^{7}$                 & $\sim$5 $\times$ 10$^{4}$ & Rarefaction wave timescale         & \cite{Tokuda_2016,Tokuda_2024} & L1521F\\
\hline
\end{tabular}
\flushleft{{\bf Table notes:} [1] Distance measurements toward subregions in Taurus by \cite{Galli_2019}. Because there is no available value for the L1521 region in their study, we adopted the representative distance to the Taurus molecular cloud, 140\,pc \cite[e.g.,][]{Elias_1978}, for MC~27.
[2] The relative age from the moment of star formation, defined as the time when the first hydrostatic core (stellar core) is formed ($t$ = 0). }
\end{table}

We conducted ALMA Cycle~10 observations targeting the three sources using the 7\,m array and total power (TP) array. The ACA observation period spanned from October 2023 to September 2024. For the main target line, ortho-H$_2$D$^+$, we allocated a central frequency of 372.42078\,GHz with a bandwidth of 62\,MHz and a frequency resolution of 61\,kHz. Additionally, continuum bands centered at 357.99950\,GHz and 369.92909\,GHz, each with a 2\,GHz bandwidth, were included. Two basebands centered at 360.16919\,GHz (frequency resolution = 30\,kHz) and 372.67200\,GHz (frequency resolution = 61\,kHz) were used to target DCO$^+$(5--4) and N$_2$H$^+$(4--3) lines, respectively.

We used the Common Astronomy Software Application package \citep{CASA_2022}, v6.6.4-34, for data reduction. The \texttt{tclean} task was performed with the multi-scale deconvolver, and clean masks were manually selected. Imaging was carried out with natural weighting to maximize sensitivity. As described in Section~\ref{sec:res}, some sources were not detected with the 7\,m array, and for those cases, mask regions were not selected. The velocity grid was set to 0.05\,km\,s$^{-1}$ for imaging. The sensitivity of the H$_2$D$^+$ data obtained with the 7\,m array was 0.1\,K, with a beam size of 4\farcs5 $\times$ 3\farcs5 ($\sim$500\,au). For the TP array, the sensitivity was 0.025\,K, and the beam size was 17\farcs5 ($\sim$2,500\,au), which corresponds to the maximum recoverable scale of the 7\,m array. In this study, the 7\,m array data and the TP array data were analyzed independently without combining the two datasets. The 0.83\,mm continuum data were obtained using only the 7\,m array. The total bandwidth was 4\,GHz, and the sensitivity was 0.78\,mJy\,beam$^{-1}$.

\section{Results} \label{sec:res}

\begin{figure}[htbp]
    \centering
    \includegraphics[width=0.9\columnwidth]{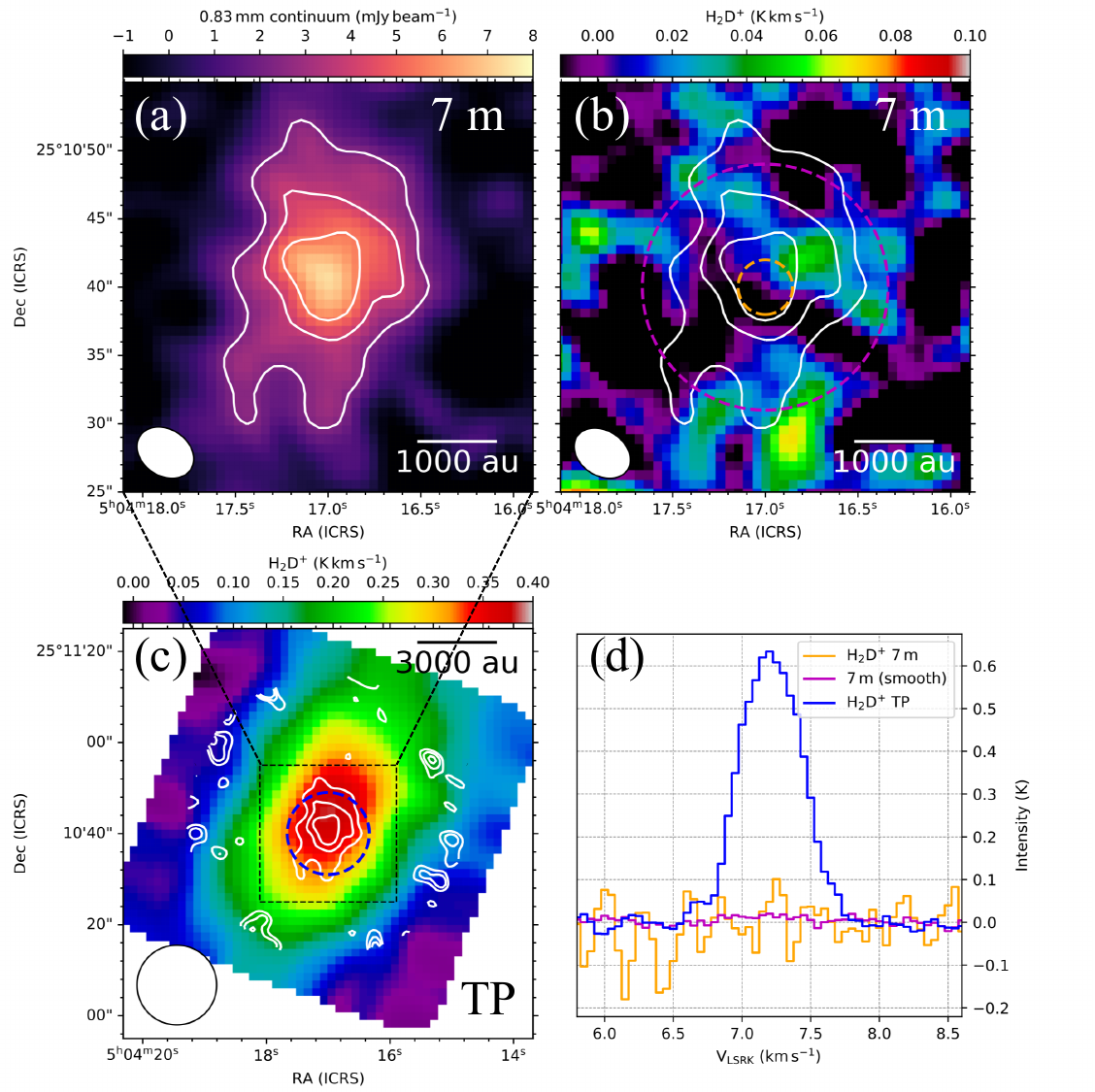}
    \caption{Spatial distribution and spectra of the 0.83\,mm continuum and ortho-H$_2$D$^+$(1$_{\rm 1,0}$--1$_{\rm 1,1}$) in L1544. {\bf (a)} The color-scale image shows the 0.83\,mm continuum obtained with the 7 m array. The contours represent the continuum emission at levels of 3, 5, and 7$\sigma$ (1$\sigma$ = 0.78 mJy\,beam$^{-1}$). The synthesized beam size is shown in the bottom-left corner. {\bf (b)} The color-scale image shows the integrated intensity of H$_2$D$^+$ over the velocity range of 7.0--7.5\,km\,s$^{-1}$ obtained with the 7 m array. The contours and the ellipse in the bottom-left corner are the same as those in panel (a). {\bf (c)} The color scale shows the integrated intensity of H$_2$D$^+$ over the velocity range of 6.8--7.8\,km\,s$^{-1}$ obtained with the TP array. The beam size of the TP array, 17\farcs5, is shown in the bottom-left corner. The contours are the same as those in panel (a).  
    {\bf (d)} The yellow, magenta, and blue spectra represent the H$_2$D$^+$ spectra obtained with the 7\,m array, the 7\,m array averaged over a region corresponding to the TP beam size, and the TP array, respectively. The extraction regions are indicated by dashed circles in the corresponding colors in panels (b) and (c).}
    \label{fig:L1544}
\end{figure}

We describe the images and spectra obtained for the three sources in 0.83\,mm continuum with the 7 m array, H$_2$D$^+$ with the 7\,m and TP arrays. Figure~\ref{fig:L1544} presents the results for L1544. The 0.83\,mm continuum image shows a compact structure of $\sim$1,000\,au, which corresponds to the one previously observed in the 1\,mm band with ALMA \citep{Caselli_2019,Furuya_2024}. The 0.87\,mm continuum intensity at the peak is measured to be 7\,mJy\,beam$^{-1}$, which corresponds to an $N_{\rm H_2}$ of $6 \times10^{22}$\,cm$^{-2}$ assuming $T_{\rm d}$ = 7\,K \citep{Evans_2001} and $\kappa_{\rm 0.8\,mm}$ = 0.01\,cm$^{2}$\,g$^{-1}$ for the previous submilletemar works \citep[e.g.,][]{Ward-Thompsoon_1994}. While the continuum emission is clearly detected, H$_2$D$^+$ is not detected above the 3$\sigma$ level in the 7\,m array data, as shown in panels (b) and (d). This non-detection does not immediately indicate the absence of H$_2$D$^+$ along the line of sight. As shown in panels (c) and (d), H$_2$D$^+$ is detected with the TP array, but even when the 7\,m array data are averaged over a region corresponding to the beam size of the TP array, no significant emission is visible. These results show that most of the H$_2$D$^+$ emission originates from structures larger than the maximum recoverable scale of the 7\,m array, resulting in it being resolved out by the interferometer. Assuming the LTE (local thermodynamic equilibrium) condition with an excitation temperature ($T_{\rm ex}$) of 7\,K \citep[c.f.,][]{Caselli_2008}, the ortho-H$_2$D$^+$ column density ($N_{\rm o-H_2D^+}$) is calculated to be $<$1 $\times$10$^{12}$\,cm$^{-2}$, using $\Delta v$ = 0.1\,km\,s$^{-1}$ (= 2 channels) and $T_{\rm b}$ = 0.3\,K as the observed upper limit. Therefore, we constrain the fractional abundance of ortho-H$_2$D$^{+}$ relative to H$_2$ within the compact structure traced by the 7\,m array continuum emission to be $\lesssim 2 \times 10^{-11}$. This estimate is derived at the continuum peak (core center) from the 0.83\,mm peak flux density (Jy\,beam$^{-1}$), evaluating the line and continuum at the same position and with the same beam, and thus refers solely to the dense, spatially compact component detectable with the interferometer. In contrast, for L1544 specifically, single-dish determinations (e.g., $\sim$10$^{-9}$; \citealt{Caselli_2003}) were derived with much lower spatial frequency coverage and are therefore sensitive to more extended emission and integration along the full line of sight. The difference in spatial sensitivity should be considered when comparing these values.

\begin{figure}[htbp]
    \centering
    \includegraphics[width=0.9\columnwidth]{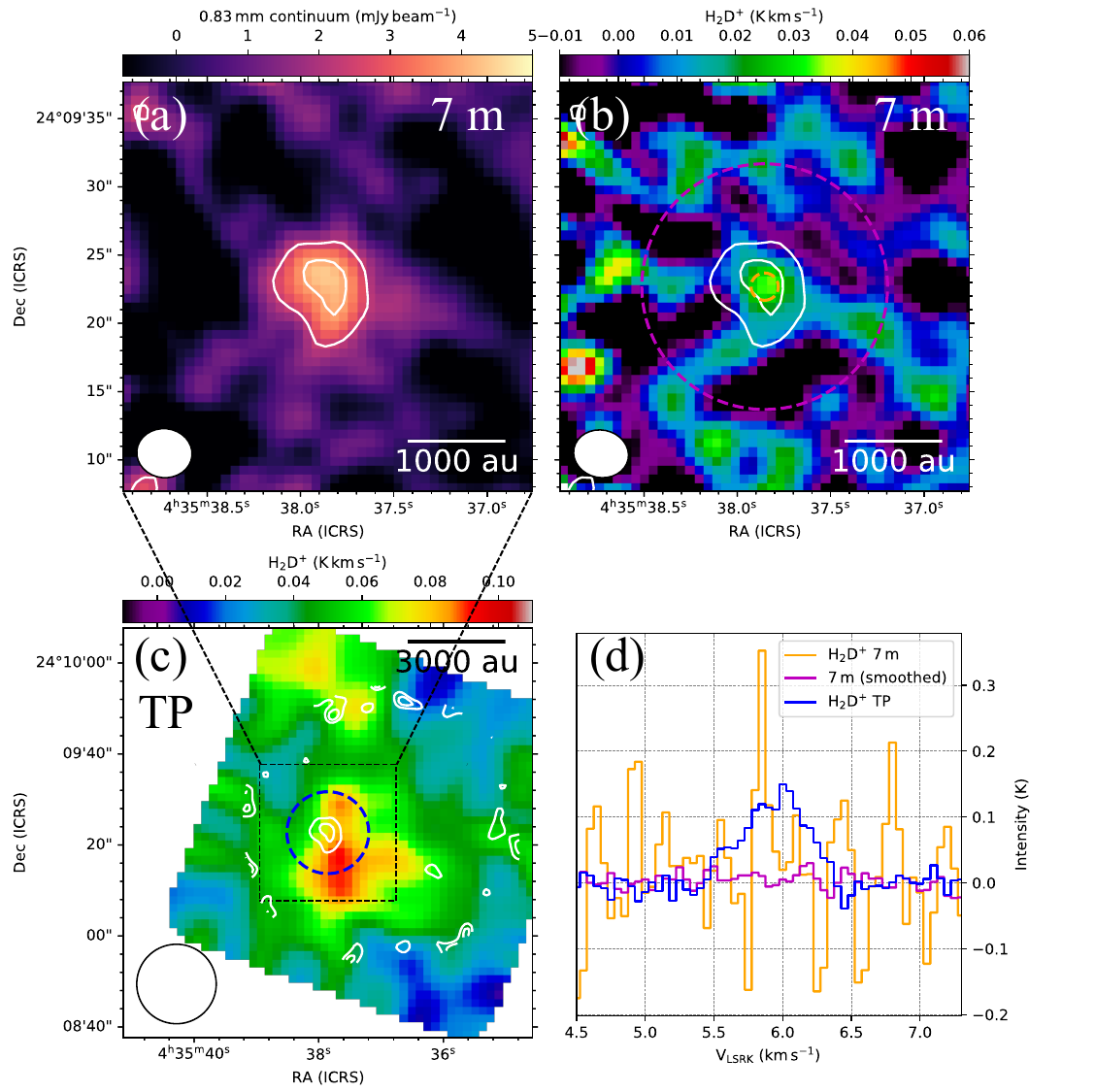}
    \caption{Same as those in Figure~\ref{fig:L1544} but for MC~35-mm. The velocity ranges integrated for the H$_2$D$^+$ data cube are 5.8--5.9\,km\,s$^{-1}$ in panel (b) and 5.5--6.5\,km s$^{-1}$ in panel (c).
    In panel (d), the spectrum from the 7\,m array (orange) was extracted by averaging over a smaller region centered on the emission peak in panel (b), in contrast to the broader areas used for the other two sources.
    }
    \label{fig:MC35}
\end{figure}

Figure~\ref{fig:MC35} shows the results for MC35-mm. The size of the continuum emission is somewhat smaller than that of L1544, and the peak intensity is weaker, $\sim$4\,mJy\,beam$^{-1}$. These spatial features are consistent with previous 1.3\,mm studies with ALMA \citep{Fujishiro_2020}. The integrated intensity map of ortho-H$_2$D$^{+}$ obtained with the 7\,m array shows faint emission at the location of the 0.83\,mm continuum source. Panel (d) displays the spectrum at this position, revealing a $\sim$0.35\,K (=3.5$\sigma$) detection. Note that we attempted spatial and spectral smoothing to improve the signal-to-noise ratio, but no significant enhancement was obtained. This suggests that, even if the emission is real, it is confined to a very narrow range both spatially and in velocity.　Using the same methodology as for L1544, and adopting the same $T_{\rm d}$ for the 0.87\,mm continuum and $T_{\rm ex}$ for ortho-H$_2$D$^+$ of 7\,K, the 7\,m array data at the peak position yield beam averaged column densities of $N_{\rm H_2}\sim4\times10^{22}$\,cm$^{-2}$ and $N_{\rm o-H_2D^+}\sim1\times10^{12}$\,cm$^{-2}$. The resulting abundance ratio of [ortho-H$_2$D$^{+}$]/[H$_2$] is $3 \times 10^{-11}$. On the other hand, the TP array clearly detects H$_2$D$^+$ with an extended spatial distribution. Unlike the elliptical shape observed in L1544, the spatial distribution in MC~35 is more complex.

\begin{figure}[htbp]
    \centering
    \includegraphics[width=0.9\columnwidth]{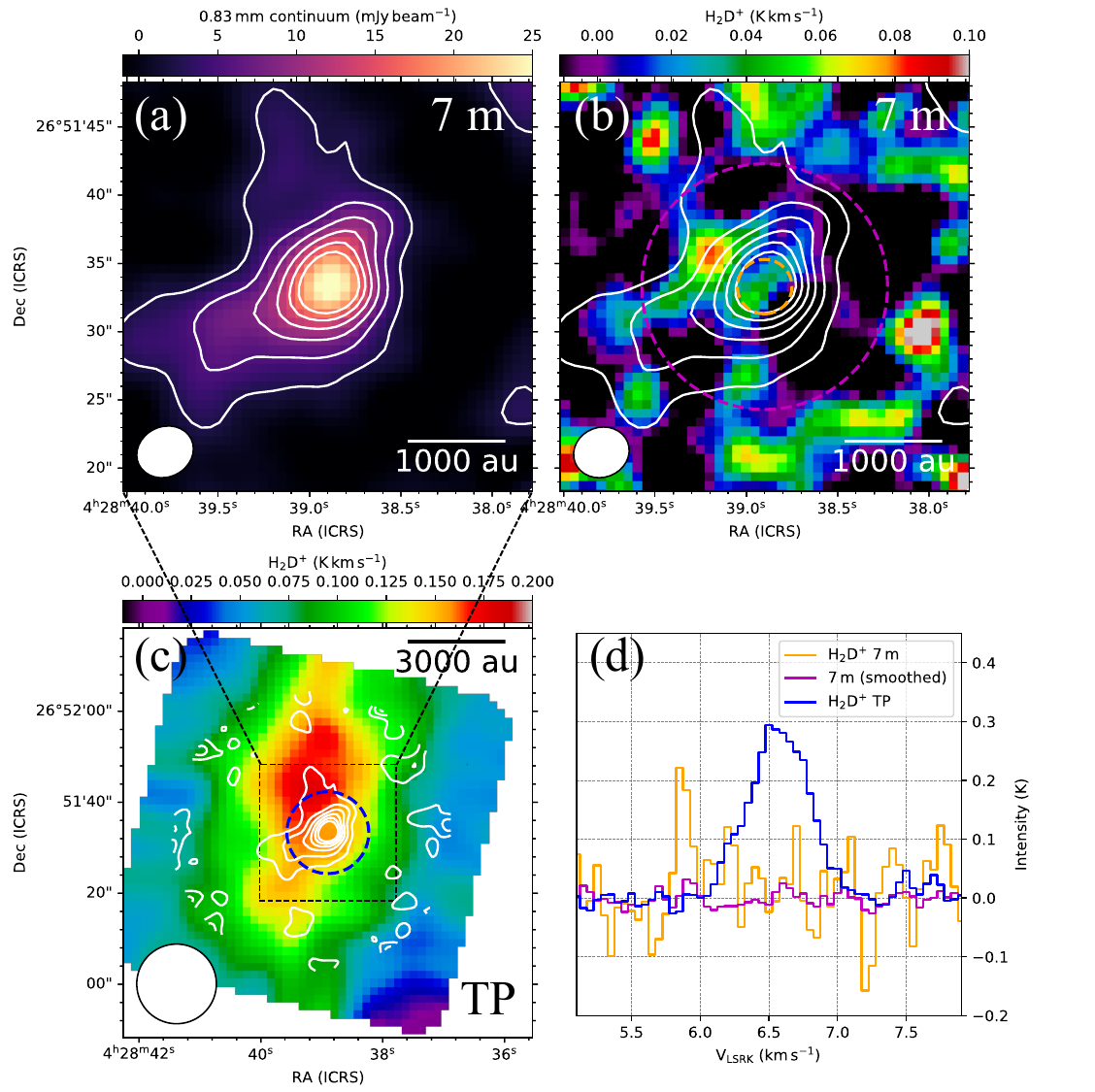}
    \caption{Same as those in Figure~\ref{fig:L1544} but for MC~27. The velocity ranges integrated for the H$_2$D$^+$ data cube are 6.0--7.0\,km\,s$^{-1}$ in panel (b) and (c).}
    \label{fig:MC27}
\end{figure}

Figure~\ref{fig:MC27} describes the results for MC~27 in the same format as the other two sources. The 0.83\,mm continuum emission (panel a) shows a spatial distribution consistent with previous studies at nearly the same wavelength \citep{Tokuda_2016}. Single-dish observations at the same frequency band showed that the present three targets have almost identical continuum intensities (see Section~\ref{sec:obs}). However, MC~27 exhibits the strongest intensity in the 7\,m array data. While this may partially result from the contribution of the faint protostellar disk \citep{Tokuda_2017,Tokuda_2024}, it is more likely that other complex structures surrounding the core (see Section~\ref{sec:obs}) are also spatially compact and detectable with interferometric observations. H$_2$D$^+$ is detected with the TP array but not with the 7\,m array (panels b, c, and d), which is the same behavior as in L1544. Although a $\sim$2$\sigma$ feature is visible at $\sim$5.8\,km\,s$^{-1}$ with the 7\,m array, it is more likely to be noise because the peak velocity is significantly different from that with the TP array (panel d). Following the same method as for the other two sources, we estimate the H$_2$ column density from the 0.83\,mm continuum intensity and derive an upper limit on the o-H$_2$D$^+$ column density. Assuming an excitation temperature of $T_{\rm ex} = 9$\,K \citep{Caselli_2008} and a line width of $\Delta v$ = 0.1\,km\,s$^{-1}$, we obtain $N$(o-H$_2$D$^+$) $\lesssim$ 6 $\times$10$^{11}$\,cm$^{-2}$. In the 7\,m array data, the 0.83\,mm continuum flux at the MC~27 core peak is measured to be 26\,mJy. The previous 12\,m array observations \citep{Tokuda_2017} estimated that $\sim$4\,mJy of this flux arises from the protostellar disk component. Subtracting this contribution, the remaining flux attributable to the envelope is $\sim$22\,mJy. Using this value with an assumpsion of $T_{\rm d}$ = 9\,K, which is the same as $T_{\rm ex}$ of ortho-H$_2$D$^{+}$ and consistent with the previous studies \cite[c.f.,][]{Tokuda_2016}, we estimate the H$_2$ column density to be $N$(H$_2$) $\sim 9 \times10^{22}$\,cm$^{-2}$. Combined with the non-detection of o-H$_2$D$^+$ in the 7\,m array, we derive an upper limit on [o-H$_2$D$^+$]/[H$_2$] within the compact structure of $\lesssim$7 $\times$10$^{-12}$.

We briefly report the detection details of other simultaneously observed lines, DCO$^+$(5--4) and N$_2$H$^+$(4--3) toward the three sources. With the TP array, all three targets show detections above the 3$\sigma$ level. In the 7\,m array data, DCO$^+$(5--4) is detected in MC35-mm and MC~27 at/around the continuum peaks, while N$_2$H$^+$ is visible only in MC~27. In L1544, none of the observed lines were detected with the 7\,m array. These results indicate that even molecular line emissions clearly detected with single-dish observations in extremely dense and evolved prestellar cores like L1544 are not necessarily detectable with interferometers.

\section{Discussion} \label{sec:dis}

Dense prestellar cores in the solar neighborhood generally exhibit large projected sizes with spatially smooth structures. Toward the moment of star formation, it is naturally expected that a centrally concentrated structure will be formed inside them. However, due to the short timescale of the high-density prestellar phase (e.g., $>$10$^6$\,cm$^{-3}$), such compact structures are rare as demonstrated by recent interferometric observation in millimeter continuum  \citep{Tokuda_2019MC5,Tokuda_2020Tau}. Therefore, understanding which molecular lines are best suited for characterizing compact high-density structures on the verge of star formation remains an ongoing challenge. In this section, we first revisit results obtained with single-dish observations in H$_2$D$^+$ and then discuss additional insights gained from our present interferometric observations. \cite{Caselli_2008} conducted a survey of ortho-H$_2$D$^+$ in various star-forming cores at the Caltech Submillimeter Observatory with a sensitivity of $\sim$0.05--0.3\,K at a velocity resolution of 0.1\,km\,s$^{-1}$. For instance, in B68, which has a central (column) density an order of magnitude lower than the targets in this study, H$_2$D$^+$ was not detected with a single-dish telescope. Similarly, in Taurus, another protostellar core containing a very-low luminosity object, IRAM~04191+1522, which is more evolved than MC~27 did not show H$_2$D$^+$ emission. These results suggest that even among dense pre- and proto-stellar sources, H$_2$D$^+$ emission is only observed in very specific and limited evolutionary stages close to the moment of star formation (see also \citealt{Koumpia_2020}). 

In our present study, H$_2$D$^+$ is detected with the 7\,m array only in the first core candidate MC~35-mm. During the prestellar phase, gravitational collapse leads to a centrally concentrated density distribution. 
Assuming a constant fractional abundance of H$_2$D$^+$ with respect to H$_2$, its emission should follow the same density evolution as traced by the continuum, and thus be detectable toward all compact cores.
However, the behavior of H$_2$D$^+$ is more complex than that in the density evolution. Because the ionization degree decreases with density (approximately $\propto n^{-1/2}$), the H$_2$D$^+$ distribution tends to be less centrally concentrated than those of the H$_2$ gas density \citep[e.g.,][]{Aikawa_2005,Aikawa_2012}. In the innermost regions, H$_2$D$^+$ is further converted into D$_2$H$^+$ and D$_3^+$, leading to a drop in its abundance. In protostellar cores, the inner density profile becomes shallower as described in the inside-out collapse model \citep{Shu_1977,Tokuda_2016}, and the rise in temperature due to protostellar heating can destroy H$_2$D$^+$. In the protostellar core MC~27, the evolved nature likely explains the non-detection of H$_2$D$^+$ with the 7\,m array. Still, extended emission in cooler outer regions remains visible in the TP data.

To better understand the behavior of H$_2$D$^+$ before protostar formation, it is useful to compare the results of L1544 and MC~35-mm. Observationally, H$_2$D$^+$ appears to be distributed so smoothly in the prestellar stage that it cannot be recovered even with the ACA 7\,m array. Considering that the continuum emission is detected in L1544, H$_2$D$^+$ should have a shallower distribution than those of $N_{\rm H_2}$. The H$_2$D$^{+}$ distribution becomes only slightly centrally concentrated in the first-core phase.

\begin{figure}[htbp]
    \centering
    \includegraphics[width=0.9\columnwidth]{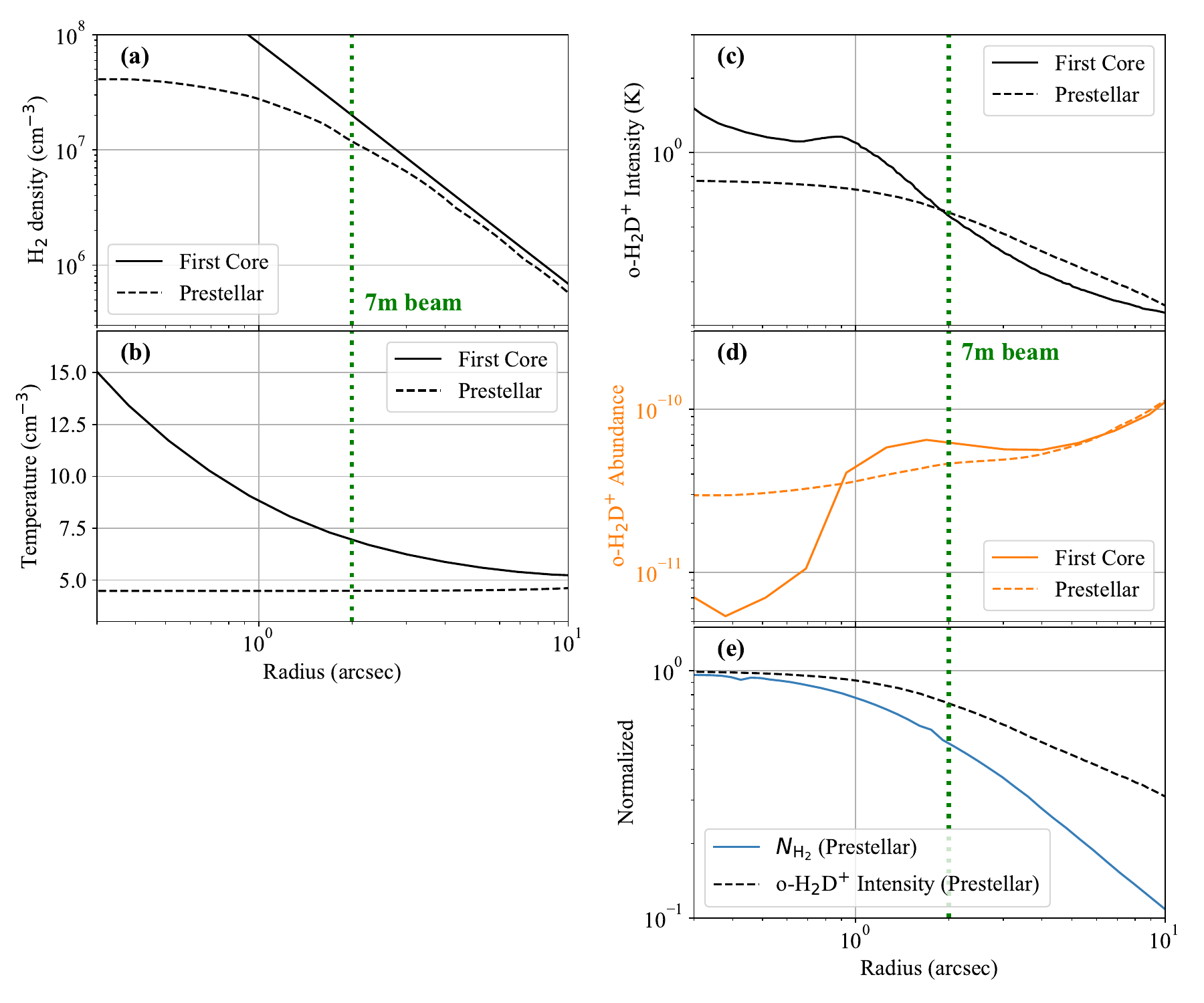}
    \caption{
    (a) H$_2$ density and (b) temperature profiles adapted from \cite{Masunaga_2000}. The $x$-axis represents the radial offset in arcseconds assuming the typical distance to the Taurus molecular cloud ($D$ $\sim$140\,pc).
    (c) Radial intensity (peak brightness temperature) profiles of ortho-H$_2$D$^+$ derived from the first-core and prestellar core models.  The green dotted line shows half of the beam size, 2$\arcsec$, with the 7\,m array. 
    (d) Radial profiles of the fractional abundance of ortho-H$_2$D$^+$ relative to H$_2$ for the first-core and prestellar core models.
    (e) The normalized radial intensity profile of ortho-H$_2$D$^+$ and $N_{\rm H_2}$ from the prestellar core model.
    }
    \label{fig:profile}
\end{figure}

To verify the above scenario, we performed chemical network calculations \citep{Furuya_2015} under the physical conditions taken from the one-dimensional core collapse model of \cite{Masunaga_2000}. We use the radial profiles of density, temperature (see Figure~4(a,b)), and $A_{\rm v}$ at two different evolutionary phases: prestellar phase ($t_{\rm core} = -6.4\times10^3$\,yr) and first hydrostatic core phase ($t_{\rm core} = -5.6\times10^2$\,yr), where $t_{\rm core} = 0$ yr is defined as the birth of the protostar. 
At each radius in the two core models, we assumed a static physical structure and calculated the chemical abundances, including o-H$_2$D$^+$, by evolving a chemical reaction network for 2\UTF{2009}$\times$10$^5$ years.
After the chemical network calculations, LTE radiative transfer simulations were performed with LIME code \citep{Brinch_2010} to compute the ortho-H$_2$D$^+$ peak brightness temperature distribution. The prestellar and first hydrostatic core phase models mimic L1544 and MC~35-mm, respectively. In Figure~\ref{fig:profile}(c), the first core model exhibits a slightly centrally concentrated intensity distribution than that of the prestellar model at a size scale of less than $\sim$2$\arcsec$, which is half of the observed beam size. This feature is consistent with the 7\,m array detection toward MC~35-mm. The abundance profile also shows a localized enhancement at a radius roughly corresponding to the beam size, but it decreases again toward the very center (see Figure~\ref{fig:profile}(d)). Panel (e) compares relative $N_{\rm H_2}$ and H$_2$D$^+$ intensity distributions in the prestellar model. It reveals that H$_2$D$^+$ has a shallower distribution than that in $N_{\rm H_2}$, explaining the (column) density-tracing continuum detection without H$_2$D$^{+}$ emission at the core center. Note that in Figure~\ref{fig:profile}(c), the absolute values of H$_2$D$^+$ intensity do not necessarily match the present observations. Our models serve as a demonstration to qualitatively understand the behavior of H$_2$D$^+$ emission during prestellar to protostellar core evolution.

To interpret the observed differences in H$_2$D$^+$ emission among the targets, we consider the underlying chemical evolution and its dependence on physical conditions such as density and temperature, as well as connections with other deuterated species. H$_2$D$^+$ emission is visible in regions with densities above 10$^5$ cm$^{-3}$, which is close to the critical density of the 372\,GHz transition \citep[e.g.,][]{Hugo_2009}, where it is spatially extended and thus detectable with the TP array. According to calculations by \cite{Walmsley_2004}, in the density range of 10$^6$--10$^7$ cm$^{-3}$, H$_2$D$^+$ is converted into more deuterated isotopes toward D$_3^+$. In fact, a single-dish work of L1544 indicates of para-D$_2$H$^+$ emission \citep{Vastel_2006}. 
This conversion leads to a sharp decrease in the fractional abundance of ortho-H$_2$D$^+$, reaching values as low as 10$^{-10}$\UTF{2013}10$^{-11}$ in the innermost region according to chemical models. These values are consistent with the upper limit derived from our 7\,m array observations of L1544, which specifically constrain the abundance within the compact central structure (scales of a few hundred to 1000\,au) traced by the interferometric continuum emission (see Section~\ref{sec:res}).
On the other hand, in the first-core phase, we suggest that a modest increase in temperature, which is not high enough to cause significant CO desorption, may enhance the H$_2$D$^+$ intensity and make detection possible with the 7\,m array. This effect is attributed to the high upper energy level ($E_{\rm u}$ = 17.9 K) from the ground state of o-H$_2$D$^+$, which leads to a substantial increase in brightness even with a small temperature rise (e.g., from 10\,K to 15\,K). Although DCO$^+$(5--4) is detected toward MC~35-mm, which could imply partial CO desorption, this interpretation remains uncertain. To further investigate the role of H$_2$D$^+$ as a pre- and proto-stellar core nucleus tracer, future studies should include comparative analyses with its daughter molecules, such as N$_2$D$^+$, and more highly deuterated species such as D$_2$H$^+$.

\section{Concluding remarks}\label{sec:Conclude}

We conducted ACA stand-alone observations with a spatial resolution of $\sim$500\,au and a maximum recoverable scale of $\sim$2,500\,au in the ortho-H$_2$D$^+$ line toward three Taurus dense cores, which are thought to be in the immediate pre- and protostellar stages. While ortho-H$_2$D$^+$ was detected in all targets with the single-dish TP array, it was only detected in MC~35-mm with the interferometer, 7\,m array. The results suggest that ortho-H$_2$D$^+$ may undergo rapid depletion in compact, high-density structures of approximately 1,000\,au. In such environments, it is likely converted into more highly deuterated molecules due to elevated densities, making ortho-H$_2$D$^+$ less effective as a tracer of core nuclei when observed with interferometers toward nearby low-mass star-forming regions. Nevertheless, our findings imply that ortho-H$_2$D$^+$ can still be a valuable tracer for identifying evolved cores just before/after star formation, making it potentially useful in first core searches.

\begin{acknowledgments}
We would like to thank the anonymous referee for useful
comments that improved the manuscript. This paper makes use of the following ALMA data: ADS/JAO. ALMA\#2023.1.00225.S. ALMA is a partnership of ESO (representing its member states), the NSF (USA), and NINS (Japan), together with the NRC (Canada), MOST, and ASIAA (Taiwan), and KASI (Republic of Korea), in cooperation with the Republic of Chile. The Joint ALMA Observatory is operated by the ESO, AUI/NRAO, and NAOJ. This work was supported by a NAOJ ALMA Scientific Research grant Nos. 2022-22B, Grants-in-Aid for Scientific Research (KAKENHI) of Japan Society for the Promotion of Science (JSPS; grant No. JP18H05440, JP20H05645, JP21H00049, JP21K13962, and JP23H00129), and Kagawa University Research Promotion Program 2025 Grant Number 25K0D015. 
\end{acknowledgments}

\software{astropy \citep{Astropy18}, CASA \citep{CASA_2022}}

\bibliographystyle{aasjournal}

\end{document}